# Effects of shocks in stellar atmosphere models on the emission line spectrum of surrounding H II regions

C. B. Kaschinski[1,2]*, Barbara Ercolano[1,2]†

[1]*Institut für Astronomie und Astrophysik der Universitäts Sternwarte München, Scheinerstr.1, D-81679 München, Germany*
[2]*Excellence Cluster Universe, Boltzmannstr. 2, 85748 Garching, Germany*

**ABSTRACT**

Emission line studies from H II regions in our own and other galaxies require tools for the inversion of line ratios into desired physical properties. These tools generally come in the form of diagnostic ratios/diagrams that are based on grids of photoionisation models. An important input to the photoionisation models is the stellar atmosphere spectrum of the ionising sources. Among a number of potentially problematic biases introduced by a great deal of unknown variables in both the stellar atmosphere and nebular models, the current omission of shocks in the calculation of the former set of models could also threaten the accuracy of the physical interpretation of emission line ratios from H II regions. Current stellar atmosphere models that are crucial inputs to the grid of photoionisation models used to generate nebular emission line diagnostic diagrams might produce significant biases due to the omission of shocks. We therefore investigate whether a new generation of photoionisation model grids, taking shocks into account, is required to compensate for the biases. We make use of the WM-Basic stellar atmosphere code, which can account for the extra energetic emission in the stellar spectral energy distribution produced by shocks, in conjunction with the photoionisation code MO-CASSIN to determine whether shocks produce significant biases in the determination of the physical parameters of the interstellar medium and/or ionising stellar parameters. We conclude that these effects are only important for stellar sources with effective temperatures lower than 30kK and in this case they yield artificially high stellar temperatures, electron temperatures and nebular ionisation states. The magnitude of the effect is also obviously dependent on the strength of the shock and is likely to be unimportant for the majority of stellar sources. Nevertheless, we find our 20kK and 30kK shock models to strongly enhance the He II $\lambda$4686 nebular emission line (next to many other lines). This result is however not strong enough to explain previously observed He II $\lambda$4686 line emission in the spectra of H II galaxies.

## 1 INTRODUCTION

Strong emission lines in the spectra of H II regions powered by hot massive stars are a powerful tool to study the interstellar medium properties in our own and other galaxies. Numerous studies use a combination of strong lines for the determination of abundance gradients in galaxies, which are a crucial constraint for chemical evolution models. The properties of the ionising stars can also be inferred through the analysis of the nebular emission spectrum, with implications on fundamental correlations, such as the relation of stellar effective temperatures and hardness of stellar spectrum and with galactocentric distance/metallicity (e.g. Campbell et al. (1988), Martín-Hernández et al. (2002), Giveon et al. (2002), Dors & Copetti (2003)). The diagnostic potential of nebular emission lines is however somewhat limited by a number of biases and uncertainties introduced by the need of relying on photoionisation models for their interpretation. In general, the inversion of line ratios to physical parameters is performed via diagnostic diagrams based on large grids of photoionisation models. Such diagrams exist for a number of important nebular diagnostics in the optical and in the mid-infrared (e.g. Morisset et al. 2004), but their accuracy is limited by a number of necessary assumptions used in the photoionisation models (e.g. gas geometry and spatial distributions of the ionising sources, Ercolano et al. (2007, 2008)) and on the stellar atmosphere models used as inputs in the photoionisation models. Morisset et al. (2004) and Simón-Díaz & Stasińska (2008) present extensive comparative studies of different stellar atmosphere codes and the effects on nebular emission line diagnostic diagrams, discussing also the error implied for the determination of gas and stellar parameters.

In addition to the well known biases and uncertainties already pointed out by the studies above, the neglect of shocks[1] in stellar atmosphere calculations presents a further threat to the interpretation of nebular emission lines via the application of model grids. The general effect of shocks in the calculation is the production of a somewhat harder stellar spectrum for the same nominal stellar effective temperature. Indeed, the importance of accounting for the effect of shocks in the calculations of stellar atmosphere wind models was already introduced by Lucy & Solomon (1970) who found that radiation driven winds are inherently unstable. The emergent X-rays from a star were then described by Lucy & White (1980) and Lucy (1982) as a radiative loss of post-shock regions where shocks are propelled by non-stationary features in the wind structure. The primary effect of these shocks in general affects the ionisation equilibrium in the stellar atmosphere with regard to the high ionisation stages like $N$ v, $O$ vi and $S$ vi. The detection of those high ionisation stages in wind spectra was referred to as the so called "superionisation" problem (cf. Snow & Morton 1976), which could be well accounted for by a first, simplified description of the shock theory. The implementation of shocks in the stellar atmosphere code WM-Basic and an application on one of the best known massive O stars $\zeta$ Puppis was presented by Pauldrach et al. (1994b) with encouraging results concerning the modeling of the $O$ vi line. During their study, Pauldrach et al. (1994b) also found that shock emission surprisingly has a non-negligible influence at radio (10%) and IR (30%) wavelengths. In their first approach, Pauldrach et al. (1994b) assumed a simplified shock description. The characterization of the shock emission is thereby mainly based on an immediate post-shock temperature which means non-stratified, isothermal shocks are considered. An improved version of WM-Basic was published by Pauldrach et al. (2001), implementing a cooling structure with a certain range of shock temperatures. This was done by adding a spatial fragmentation of the shock structure. The inner region of the stellar wind consists of radiative shocks (cf. Chevalier & Imamura 1982) where the cooling time is shorter than the flow time. In the outer region on the other hand, the velocity structure reaches its stationary value, i.e. the terminal velocity. This leads to a negligible radiative acceleration and hence a large flow time. The cooling processes can thus be approximated by adiabatic expansion (Simon & Axford 1966). An already modified approach of isothermal wind shocks by Feldmeier et al. (1997) was used for implementation in WM-Basic. The improved code yielded again almost perfect fits of the highly ionised elements in the observed spectrum of $\zeta$ Puppis. The improved code yields a much more efficient radiation of shocks in the soft X-ray spectral band. A comparison to the observed ROSAT PSPC spectrum showed fits of at least the same quality as those obtained by Feldmeier et al. (1997). During the same study Pauldrach et al. (2001) also presented an analysis of the supergiant $\alpha$ Cam which was based on their improved shock method. The resulting model matched the UV observations almost perfectly. Further observational investigations regarding the detection of X-rays from O stars were performed by Sana et al. (2005) and Sana et al. (2006) for the rich O-type star cluster NGC 6231 and by Albacete Colombo et al. (2003) for the Carina nebula field. The observational data for the two studies have been both obtained by the XMM-Newton satellite. The observations revealed several hundred X-ray sources[2] for the NGC 6231 cluster and 80 discrete ones for the Carina nebula.

* E-mail:corni@usm.uni-muenchen.de
† E-mail:ercolano@usm.uni-muenchen.de

The shock enhanced emission of the stellar atmosphere models is also relevant for the ionisation of a number of abundant species in nebular gas surrounding hot stars. It is therefore important to ask the question of whether the current neglect of shocks in the stellar atmosphere models used as inputs to the grid of photoionisation models used by nebular emission line diagnostic diagrams would produce significant biases, and therefore whether a new generation of photoionisation model grids should be calculated.

In this paper we use the stellar atmosphere code WM-Basic (Pauldrach et al. 2001) and the photoionisation code MOCASSIN (Ercolano et al. 2003, 2004, 2005) to investigate the effects of shock-enhanced stellar spectral energy distributions on the emission line spectra of H ii regions. In particular we investigate the bias introduced on the determination of gas ionisation parameters, metallicities and stellar effective temperatures from common nebular emission line diagnostics. While a subset of the stellar atmosphere calculations are discussed in this work, we have also calculated a bigger library which includes a larger range of stellar temperatures. This is provided online for use with the photoionisation code MOCASSIN.

A brief description of the methods employed in this paper, including a description of the numerical codes is given in Section 2. The effects on nebular diagnostics as calculated by the photoionisation models are discussed in Section 3. Section 4 consists of a brief summary of our results and main conclusions.

## 2   METHODS

The unstable, non-stationary behavior of stellar atmosphere winds produces shock cooling zones. As a consequence, these cooling zones radiate X-rays which can be observed. This has been studied for a variety of objects. The massive O star $\zeta$ Puppis and the supergiant $\alpha$ Cam have been investigated by Pauldrach et al. (2001) who obtained values for the total X-ray luminosity of $log(L_X/L_{bol}) = -7.1$ and $log(L_X/L_{bol}) = -6.5$, respectively. Sana et al. (2005) and Sana et al. (2006) presented an analysis of the young open cluster NGC 6231 and derived for the contained O stars an X-ray luminosity value of $log(L_X/L_{bol}) \approx -7.0$. They also reviewed the Carina nebula field observations presented by Albacete Colombo et al. (2003). On the bases of their analysis, Sana et al. (2006) rederived for the observed O stars a value of $log(L_X/L_{bol}) \approx -6.2$ for the integrated X-ray luminosity. For our computed models we have therefore chosen similar parameter values which are thus representative of the real situations. We have run a grid of 16 stellar atmosphere models using the WM-Basic code in shock and no-shock mode for stellar effective temperatures in the range of 20kK to 50kK in steps of 10kK for a surface gravity of log g = 3.9 and a temperature range of 60kK to 90kK also in steps of 10kK for a surface gravity of log g = 4.9. All models, i.e. shock as well as no-shock, use solar abundances. The parameters for the shock models are:

- $v_t/v_\infty$: The jump velocity $v_t$ is the immediate post-shock temperature normalized to the terminal velocity $v_\infty$. $v_t$ was set to 0.25 in units of the terminal velocity $v_\infty$.
- *X-ray luminosity*: The X-ray luminosity is normalized to the bolometric luminosity ($log(L_X/L_{bol})$). We used a value of $-6.5$.

---

[1] Shocks are understood as EUV and X-ray radiation produced by cooling zones that are based on shock heated matter which in turn is the result of the non-stationary, unstable behavior of radiation driven winds.

[2] A smaller fraction of which are O stars.

- *Parameter m*: The parameter m controls the onset of shocks. It is defined as the ratio of outflow to sound velocity where shocks start to form. A direct correlation of jump velocity to outflow velocity reflects a dependency to the radius. This means via this parameter it can be controlled where shocks start to form in the wind, e.g. higher values for m enables shocks for outer regions but leaves inner regions mainly undistured. m = 1 was adopted enabling shocks in the whole atmosphere.

- *Stratification exponent $\gamma$*: $\gamma$ controls the strength of the shocks depending on the local velocity $v(r)$. It is set to 1 for all simulations, which means a linear dependency on the local velocity.

The models were formatted for easy use in the MOCASSIN code and are freely downloadable from www.3d-mocassin.net. For our grid of effective temperatures and surface gravities a radius of 2.2 solar radii $R_\odot$ was assumed for all shock and no-shock models. The models were used as input to the MOCASSIN photoionisation code. The photoionisation models were set up as to simplify the configuration as much as possible in order to easily isolate the effects of shocks on the physical conditions in the nebular gas. The set up is very similar to that employed in the Meudon-Lexington benchmark models (Péquignot et al. 2001) and assume spherical geometry and homogeneous gas distributions. The input ionising flux is set to the output of the stellar atmosphere calculations and is given in Table 1. This corresponds to models with mean ionisation parameters log (U) = -2.6, -2.4. and -2.2 respectively for $T_{eff}$ = 20, 30 and 40kK. The outer radius of the photoionisation models were adjusted to the nebular Strömgren radius. The set up parameters are as follows:

- *Formatted WM-Basic spectrum*: Emergent synthetic spectrum (Flux of the star [$\mathring{A}$]) up to a wavelength of $1 \times 10^5$ $\mathring{A}$.
- *$T_{eff}$*: The effective stellar temperature of the WM-Basic model in Kelvin.
- *$L_{phot}$*: The number of ionising photons of the central star computed in a range up to a wavelength of $1 \times 10^5$ $\mathring{A}$.

A list of all the MOCASSIN model input parameter values for the shock and no-shock models in the temperature range of 20kK to 40kK is shown in Table 1. The models with higher temperatures than 40kK of our calculated grid are not listed since the effects of shocks are only of significance for low temperature stars. All models use a hydrogen density of 3000 [$cm^{-3}$].

## 2.1 WM-Basic

In this manuscript we will only give a short overview of the general concept and computational steps which are performed by WM-Basic. A more detailed description of the concept and all its contained fundamental methods are shown in the corresponding code publications by Pauldrach et al. (1994a), Pauldrach et al. (1994b), Pauldrach et al. (1998) and Pauldrach et al. (2001).

WM-Basic provides the tool to construct detailed atmospheric models and synthetic spectra for hot luminous stars. The concept is based on homogeneous, stationary and spherically symmetric radiation driven winds. The driving mechanism of the expanding atmospheres lies in the line absorption and scattering of Doppler-shifted metal lines (Lucy & Solomon (1970)). The model code is constituted by three main blocks:

(i) *The hydrodynamics*: In a first step the hydrodynamics is solved. The input values are the effective temperature $T_{eff}$, the logarithm of the photospheric gravitational acceleration *log g*, the photospheric radius $R_*$ and the abundances Z in units of the correspond-ing solar values. For each line the oscillator strength $f_{lu}$, the statistical weights $g_l$, $g_u$ and the occupation numbers $n_l$, $n_u$ of the lower and upper levers have to be computed. The hydrodynamics is then solved by iterating the complete continuum force $g_c(r)$, the temperature structure (assuming in both cases a spherical grey model), the density structure $\rho(r)$ and the velocity structure $v(r)$.

(ii) *NLTE-model*: During this step the radiation field $H_v(r) = 4\pi F_v(r)$, the mean intensity $J_v(r)$ and the final NLTE temperature structure $T(r)$ (determined by the energy equation and line blanketing effect which reflects the influence of line blocking on the temperature structure) are calculated. All opacities $\chi_v$ and emissivities $\eta_v$ using detailed atomic models are computed along with the occupation numbers for all relevant levels. Solving the spherical-transfer equation correctly for all total opacities ($\kappa_v$) and source functions ($S_v$) yields the radiation field. A revised treatment of shocks is included and expressed by the shock source function ($S_v^S$).

(iii) *Solution of the formal integral and computation of the synthetic spectrum*: in a final step the synthetic spectrum is computed. This is done by performing a formal integral solution of the transfer equation in the observer's frame. The synthetic spectrum is then compared to the observations, if available.

The three blocks form a coupled system and therefore an iterative solution is required. Next to a synthetic model spectrum, the ionising fluxes and a hydrodynamical structure of the wind are calculated enabling the derivation of the mass-loss rate $\dot{M}$ and the velocity field $v(r)$ which characterize the wind features.

## 2.2 MOCASSIN

The 3-dimensional Monte Carlo photoionisation code Mocassin (version 2.02.67), employed for the calculation of the nebular electron temperature, ionisation structure and emission line spectrum, is described in detail by Ercolano et al. (2003, 2005, 2008). The basic structure of the code is summarised briefly here. MOCASSIN employs a Monte Carlo approach to the transfer of radiation allowing the treatment of primary and secondary diffuse fields exactly for arbitrary geometries and density distributions. The code includes all the major microphysical processes that influence the gas ionisation balance and the thermal balance of dust and gas, including processes that couple the gas and dust phases. In the case of H $\textsc{ii}$ regions ionised by OB stars the dominant heating process for typical gas abundances is hydrogen photoionisation, balanced by cooling by collisionally excited (generally forbidden) line emission (dominant), recombination line emission, free-bound and free-free continuum emission. The atomic database includes opacities from Verner et al. (1993) and Verner & Yakovlev (1995), energy levels, collision strengths and transition probabilities from Version 5.2 of the CHIANTI database (Landi et al. 2006, and references therein) and hydrogen and helium free-bound continuous emission data of Ercolano & Storey (2006). Arbitrary ionising spectra can be used as well as multiple ionisation sources whose ionised volumes may or may not overlap, with the overlap region being self-consistently treated by the code.

## 3 RESULTS

We compared the resulting lines for our no-shock to shock models for the stellar atmosphere models of temperatures ranging from 20kK to 50kK in steps of 10kK. Tables 2 and 3 show all lines of

| Input Parameter | 20 [$10^3 K$] | | 30 [$10^3 K$] | | 40 [$10^3 K$] | |
|---|---|---|---|---|---|---|
| | ns | s | ns | s | ns | s |
| Temperature [K] | 20000 | | 30000 | | 40000 | |
| log(U) | -2.6 | | -2.4 | | -2.2 | |
| $Q_{13.6}$ | $3.0 \times 10^{47}$ | $3.0 \times 10^{47}$ | $1.2 \times 10^{48}$ | $1.2 \times 10^{48}$ | $3.0 \times 10^{48}$ | $3.0 \times 10^{48}$ |
| $r_{stroem}$ [cm] | $3.1 \times 10^{17}$ | $3.1 \times 10^{17}$ | $5.0 \times 10^{17}$ | $5.0 \times 10^{17}$ | $6.7 \times 10^{17}$ | $6.7 \times 10^{17}$ |

**Table 1.** MOCASSIN input parameter values. $r_{stroemgren}$ is the Strömgren radius and $Q_{13.6}$ the number of ionising photons of the model. Both quantities have been calculated using the ionising flux of the corresponding WM-Basic model up to a wavelength of $1 \times 10^5$ Å. The ionisiation parameters log(U) (see section 3) have been calculated according to the corresponding temperatures, ionising photons $Q_{13.6}$ and Strömgren radii $r_{stroem}$ of each model.

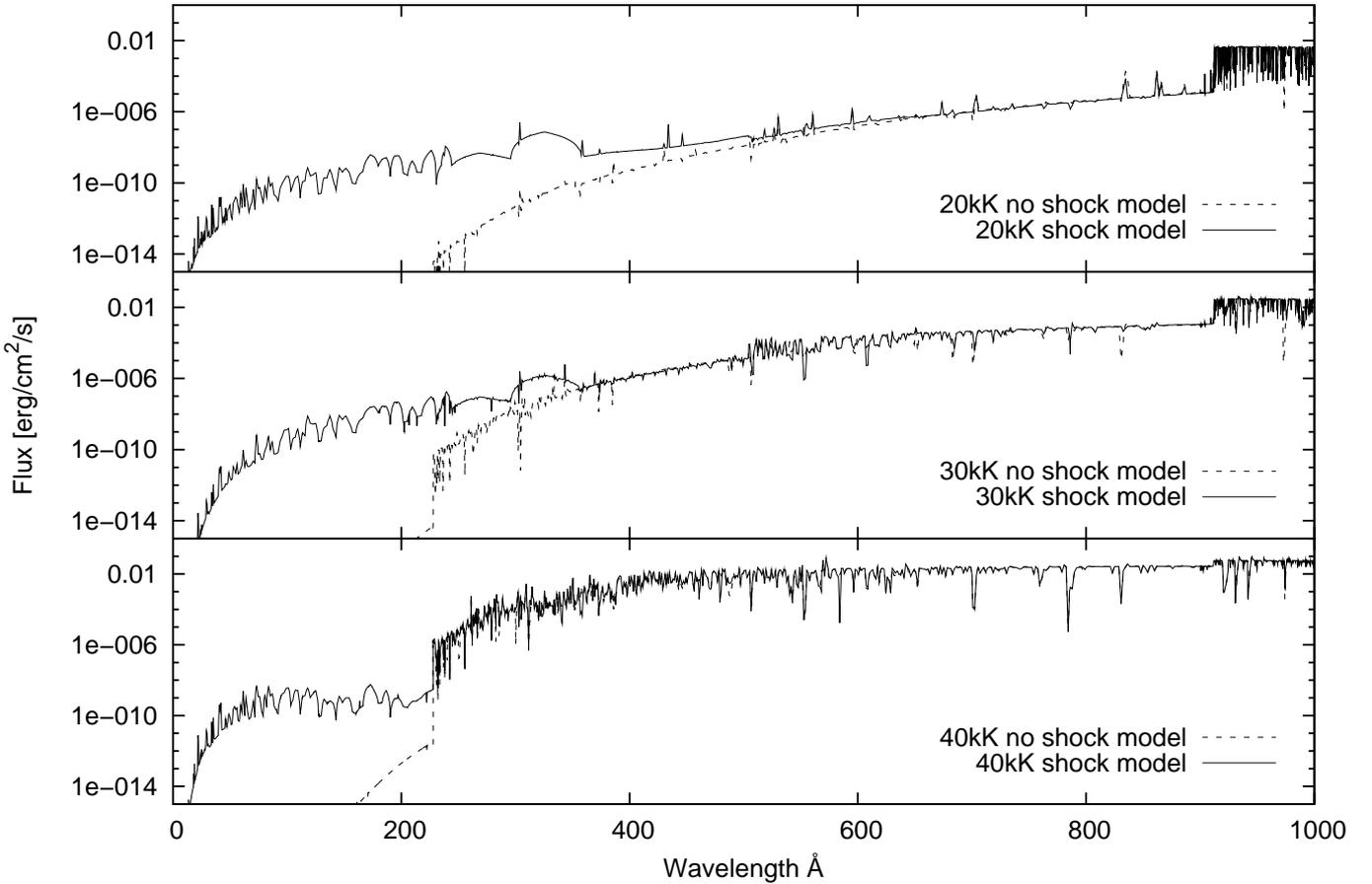

**Figure 1.** Ionising flux of our three models ranging in the effective temperatures from 20kK (top panel) to 40kK (bottom panel). Each panel shows one of the three models, computed once including shocks (solid line) and once neglecting shocks (dashed line). The comparison shows the non-negligible influence of shocks on the ionising flux.

He I, He II and the heavier elements of these runs where the differences in the line strengthes between the no-shock and shock runs are larger then 20%. As can be seen from the table only models with a temperature of 20kK and 30kK show differences. Models for higher temperatures do only show differences between no-shock to shock models which are smaller than 20% and are therefore not listed. In what follows, we used these lines to deduce physical properties of the ionising stellar source and of the nebular gas.

### 3.1  Effects of shocks on the stellar spectral energy distribution

Figure 1, compares shock (solid line) and no-shock models (dashed line) for stellar atmosphere models of temperatures 20kK (upper panel), 30kK (middle panel) and a 40kK (lower panel) with a log g of 3.9. Shock parameters are as discussed in section 2.

As shown in Fig. 1, shocks produce extra emission in the range short of approximately 500 Å for all models. However, stars with temperatures in excess of 30kK already radiate a significant fraction of their bolometric luminosity in this range thus diminishing

| Line [$\lambda$] | 20 [$10^3 K$] | | 30 [$10^3 K$] | |
|---|---|---|---|---|
| | ns | s | ns | s |
| [He $\textsc{i}$] 4027 Å | $6.741 \times 10^{-5}$ | $1.020 \times 10^{-3}$ | - | - |
| [He $\textsc{i}$] 5017 Å | $7.453 \times 10^{-5}$ | $1.136 \times 10^{-3}$ | - | - |
| [He $\textsc{i}$] 1.08 $\mu m$ | $1.072 \times 10^{-3}$ | 0.017 | - | - |
| | | | | |
| [He $\textsc{ii}$] 1215 Å | $< 1 \times 10^{-5}$ | $9.554 \times 10^{-4}$ | - | - |
| [He $\textsc{ii}$] 1640 Å | $< 1 \times 10^{-5}$ | $3.160 \times 10^{-3}$ | $< 1 \times 10^{-5}$ | $2.380 \times 10^{-4}$ |
| [He $\textsc{ii}$] 4686 Å | $< 1 \times 10^{-5}$ | $4.880 \times 10^{-4}$ | $< 1 \times 10^{-5}$ | $3.676 \times 10^{-5}$ |

**Table 2.** He $\textsc{i}$ and He $\textsc{ii}$ nebular line strengths (with respect to Hβ) corresponding to the stellar source models with effective temperatures of 20kK and 30kK. Both models have been calculated in no-shock 'ns' and shock 's' mode. Only lines which show differences larger than 20% are shown. The dash '-' appearing for some lines means differences in the line strength which are smaller than 20% for this particular line and are not listed. For stellar atmosphere models with a temperature of 30kK only the He $\textsc{ii}$ $\lambda1640$ and He $\textsc{ii}$ $\lambda4686$ lines show a large enough difference in the line strength between the no-shock and shock model. There are no differences in the line strength larger than 20% for models with higher temperatures than 30kK which are therefore not listed.

the effects of shocks. For hotter stars the shock effects are completely washed away.

The computed spectral energy distribution of stars with temperatures of 20kK is however visibly affected by the inclusion of shocks, providing an excess of photons at energies able to ionise some common ionic species found in H $\textsc{ii}$ regions. As a consequence, for these lower temperature stars, one would expect a different nebular ionisation structure resulting for photoionisation models that employ shock stellar atmosphere models as energy sources. This would also be reflected in the nebular emission line spectrum from the different models as will be discussed in the next section.

### 3.2 Implications for nebular line diagnostics of stellar effective temperature

The ionisation parameter of an ionised nebular gas can be defined as $U = Q_{13.6}/(n_H \cdot 4\pi r^2 c)$ (where $Q_{13.6}$ is the number of ionising photons per second produced by the central star, $n_H$ is the hydrogen number density, r is the Strömgren radius and c is the speed of light) and effectively provides a measure for the number of photons that are available to ionise a given pocket of gas at a given location in the nebula. Evans & Dopita (1985) define a mean $\overline{U} = U(\overline{r})$ where $\overline{r}$ is a mean distance from the ionising source. This quantity can be useful to get a handle on the level of ionisation of the gas. A number of empirical methods have been developed to estimate $\overline{U}$ and $T_{eff}$ from ratios of strong nebular emission lines, and one of the most commonly employed ones is based on a combination of ratios of the infrared lines [S $\textsc{iv}$]$\lambda10.5\mu m$/[S $\textsc{iii}$]$\lambda18.7\mu m$ and the [Ne $\textsc{iii}$]$\lambda15.5\mu m$/[Ne $\textsc{ii}$]$\lambda12.8\mu m$, via the $\eta_{S-Ne}$ parameter as used by Morisset et al. (2004) based on the original definition by Vilchez & Pagel (1988). The $\eta_{S-Ne}$ parameter is defined as

$$\eta_{S-Ne} = [S \textsc{iv}/\textsc{iii}]/[Ne \textsc{iii}/\textsc{ii}] \qquad (1)$$

where [S $\textsc{iv}$/$\textsc{iii}$] =[S $\textsc{iv}$]$\lambda10.5\mu m$/[S $\textsc{iii}$]$\lambda18.7\mu m$ and [Ne $\textsc{iii}$/$\textsc{ii}$] = [Ne $\textsc{iii}$]$\lambda15.5\mu m$/[Ne $\textsc{ii}$]$\lambda12.8\mu m$.

Table 4 shows the ratios computed by our photoionisation models for input stellar atmospheres with and without shocks labeled 's' and 'ns', respectively. These ratios and the resulting $\eta_{S-Ne}$ parameter are then plotted in Figure 2 and compared with the grid of photoionisation models presented by Morisset et al. (2004) (black solid and dashed lines), for log(U) = -1, -2 and -3 at stellar effective temperatures $T_{eff}$ = 35kK (circles), 40kK (triangles) and 45kK (diamonds).

Before we move on to the interpretation of the effect of shocks on these diagnostics it is perhaps worth noticing here that the difference between the location of our non-shock 40kK point from that of the Morisset et al. (2004) model with similar ionisation parameter is understandable due to the different input parameters used in this work for the stellar atmosphere code and for the photoionisation code. We stress that the aim here is not to benchmark our models to this previous work, rather the Morisset et al. (2004) tracks are only shown in order to illustrate the direction in which changes in the ionisation parameter and effective temperature affect the diagnostic diagram.

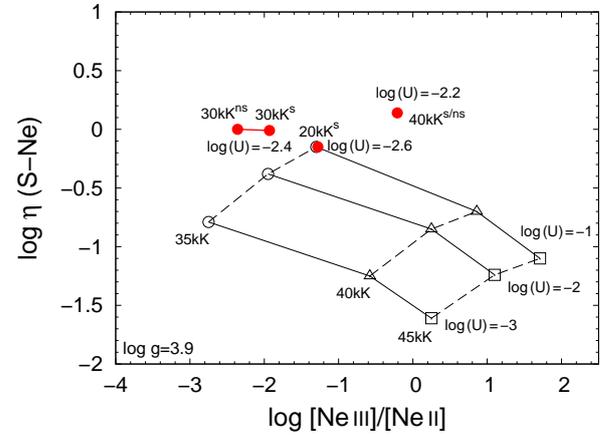

**Figure 2.** Resulting $\eta_{S-Ne}$ parameter derived for the MOCCASIN models compared to the grid of photoionisation models presented by Morisset et al. (2004). Shown are the position of the corresponding parameter values for the shock and no-shock models labeled by the 's' and 'ns' exponent. Obviously there are no differences between the 40kK shock and no-shock model. The 30kK shock model compared to the 30kK no-shock model shows already small deviations. The 20kK no-shock model is not even depicted in the plot since the spectrum of a 20kK star is not hot enough to produce a significant population of doubly ionised neon or triply ionised sulfur. The 20kK shock model is located in a region expected to be populated by systems with roughly 30-35kK ionising stars.

The net effect of shocks is to simulate a star of higher effective temperature and a higher ionisation parameter in the nebular gas as shown in Figure 2. These effects are however only important for cooler stars, indeed the 30kK shock case only shows a small deviation from the non-shock case, and no change at all is visible for the 40kK case. The effects on the 20kK case are, on the other hand, dramatic. The 20kK non-shock case is in fact not even shown on the diagram, since the spectrum of a 20kK star is not hot enough to produce a significant population of doubly ionised neon or triply ionised sulfur. The addition of shocks, however, does produce ionising photons in the required wavelength range, placing the 20kK shock model in a region of the diagnostic diagram expected to be populated by systems with roughly 30-35kK ionising stars and approximately a decade higher ionisation parameter.

| Line [$\lambda$] | 20 [$10^3 K$] | | 30 [$10^3 K$] | |
|---|---|---|---|---|
| | ns | s | ns | s |
| [C II] 2325 Å | $4.654 \times 10^{-4}$ | $5.969 \times 10^{-4}$ | $2.337 \times 10^{-3}$ | $2.318 \times 10^{-3}$ |
| [C II] 2324 Å | $4.924 \times 10^{-4}$ | $6.315 \times 10^{-4}$ | $2.337 \times 10^{-3}$ | $2.318 \times 10^{-3}$ |
| [C II] 2329 Å | $5.504 \times 10^{-4}$ | $7.059 \times 10^{-4}$ | $2.337 \times 10^{-3}$ | $2.318 \times 10^{-3}$ |
| [C II] 2328 Å | $1.507 \times 10^{-3}$ | $1.932 \times 10^{-3}$ | $2.337 \times 10^{-3}$ | $2.318 \times 10^{-3}$ |
| [C II] 2326 Å | $2.969 \times 10^{-3}$ | $3.810 \times 10^{-3}$ | $2.337 \times 10^{-3}$ | $2.318 \times 10^{-3}$ |
| [C III] 1909 Å | $< 1 \times 10^{-5}$ | $1.742 \times 10^{-4}$ | $3.210 \times 10^{-3}$ | $2.437 \times 10^{-3}$ |
| [C III] 1907 Å | $< 1 \times 10^{-5}$ | $2.277 \times 10^{-5}$ | $2.422 \times 10^{-3}$ | $1.842 \times 10^{-3}$ |
| | | | | |
| [N I] 5199 Å | $3.745 \times 10^{-5}$ | $5.843 \times 10^{-5}$ | - | - |
| [N I] 5202 Å | $1.532 \times 10^{-5}$ | $2.403 \times 10^{-5}$ | - | - |
| [N II] 2140 Å | $7.093 \times 10^{-5}$ | $9.330 \times 10^{-5}$ | - | - |
| [N II] 2143 Å | $1.749 \times 10^{-4}$ | $2.300 \times 10^{-4}$ | - | - |
| [N III] 1750 Å | - | - | $1.857 \times 10^{-5}$ | $1.290 \times 10^{-5}$ |
| [N III] 57.3 $\mu m$ | $< 1 \times 10^{-5}$ | $4.020 \times 10^{-3}$ | | |
| | | | | |
| [O I] 6302 Å | $4.324 \times 10^{-3}$ | $6.569 \times 10^{-3}$ | - | - |
| [O I] 6366 Å | $1.383 \times 10^{-3}$ | $2.101 \times 10^{-3}$ | - | - |
| [O I] 5579 Å | $1.088 \times 10^{-5}$ | $2.200 \times 10^{-5}$ | - | - |
| [O III] 88.3 $\mu m$ | $< 1 \times 10^{-5}$ | $6.919 \times 10^{-3}$ | $2.109 \times 10^{-3}$ | $4.980 \times 10^{-3}$ |
| [O III] 51.8 $\mu m$ | $< 1 \times 10^{-5}$ | $0.039$ | $0.011$ | $0.027$ |
| [O III] 4960 Å | $< 1 \times 10^{-5}$ | $0.012$ | $9.150 \times 10^{-3}$ | $0.019$ |
| [O III] 2322 Å | - | - | $< 1 \times 10^{-5}$ | $1.581 \times 10^{-5}$ |
| [O III] 5008 Å | $< 1 \times 10^{-5}$ | $3.500$ | $0.027$ | $0.056$ |
| [O III] 1666 Å | - | - | $< 1 \times 10^{-5}$ | $1.296 \times 10^{-5}$ |
| [O III] 4363 Å | $< 1 \times 10^{-5}$ | $1.592 \times 10^{-5}$ | $3.623 \times 10^{-5}$ | $6.283 \times 10^{-5}$ |
| [O IV] 25.9 $\mu m$ | $< 1 \times 10^{-5}$ | $2.564 \times 10^{-3}$ | $< 1 \times 10^{-5}$ | $2.274 \times 10^{-4}$ |
| | | | | |
| [Ne II] 12.8 $\mu m$ | $0.497$ | $0.616$ | - | - |
| [Ne III] 15.5 $\mu m$ | $< 1 \times 10^{-5}$ | $0.032$ | $3.664 \times 10^{-3}$ | $9.804 \times 10^{-3}$ |
| [Ne III] 3870 Å | $< 1 \times 10^{-5}$ | $8.971 \times 10^{-4}$ | $3.651 \times 10^{-4}$ | $8.213 \times 10^{-4}$ |
| [Ne III] 36.0 $\mu m$ | $< 1 \times 10^{-5}$ | $2.566 \times 10^{-3}$ | $3.004 \times 10^{-4}$ | $8.021 \times 10^{-4}$ |
| [Ne III] 3969 Å | $< 1 \times 10^{-5}$ | $2.703 \times 10^{-4}$ | $1.100 \times 10^{-4}$ | $2.474 \times 10^{-4}$ |
| | | | | |
| [Mg IV] 4.5 $\mu m$ | - | - | $< 1 \times 10^{-5}$ | $1.737 \times 10^{-5}$ |
| | | | | |
| [Si II] 2335 Å | $8.767 \times 10^{-5}$ | $1.276 \times 10^{-4}$ | - | - |
| [Si II] 2351 Å | $5.742 \times 10^{-5}$ | $8.355 \times 10^{-5}$ | - | - |
| [Si II] 2345 Å | $2.642 \times 10^{-4}$ | $3.836 \times 10^{-4}$ | - | - |
| [Si II] 2335 Å | $3.417 \times 10^{-4}$ | $4.942 \times 10^{-4}$ | - | - |
| | | | | |
| [S III] 33.5 $\mu m$ | $0.085$ | $0.147$ | - | - |
| [S III] 12.0 $\mu m$ | $< 1 \times 10^{-5}$ | $1.152 \times 10^{-5}$ | - | - |
| [S III] 8831 Å | $1.264 \times 10^{-4}$ | $2.290 \times 10^{-4}$ | - | - |
| [S III] 18.7 $\mu m$ | $0.160$ | $0.277$ | - | - |
| [S III] 9070 Å | $0.015$ | $0.027$ | - | - |
| [S III] 3722 Å | $1.121 \times 10^{-4}$ | $2.287 \times 10^{-4}$ | - | - |
| [S III] 9532 Å | $0.085$ | $0.153$ | - | - |
| [S III] 6314 Å | $1.982 \times 10^{-4}$ | $4.041 \times 10^{-4}$ | - | - |
| [S IV] 10.5 $\mu m$ | $< 1 \times 10^{-5}$ | $0.035$ | $4.031 \times 10^{-3}$ | $0.012$ |

**Table 3.** Resulting nebular line strengths (with respect to Hβ) for all higher elements of the stellar source models with effective temperatures of 20kK and 30kK. Both models have been calculated in no-shock 'ns' and shock 's' mode. Only lines which show differences larger than 20% are shown. The dash '-' appearing for some lines means differences in the line strength which are smaller than 20% for this particular line and are not listed. This is also true for all lines of stellar atmosphere models for temperatures higher than 30kK which are therefore also not shown.

| Ratio | 20 [$10^3 K$] | | 30 [$10^3 K$] | |
|---|---|---|---|---|
| | ns | s | ns | s |
| log([Ne III]/[Ne II]) | - | $-1.286$ | $-2.365$ | $-1.935$ |
| log([S IV]/[S III]) | - | $-0.896$ | $-2.375$ | $-1.881$ |
| log $\eta$ | - | $-0.157$ | $0.002$ | $-0.012$ |

**Table 4.** MOCASSIN logarithmic ion ratios.

### 3.3 Implications for line diagnostics of electron temperature and density

The energy-level structure of some ions leads to the formation of emission lines from two different upper levels which have considerably different excitation energies. The relative rates of excitation to these levels depend strongly on the temperature and thus emission lines originating from them may be used to measure the electron temperature. We used the line ratios of the four lines [O III], [N II], [Ne III] and [S III] to deduce the electron temperatures for the no-shock and shock runs of our 20kK, 30kK and 40kK models. The line ratios are connected to the electron temperature via the following equations (see Osterbrock & Ferland 2006):

$$[O \, \text{III}] \frac{j_{\lambda4959} + j_{\lambda5007}}{j_{\lambda4363}} = \frac{7.90 exp(3.29 \times 10^4/T)}{1 + 4.5 \times 10^{-4} n_e/\sqrt{T}} \quad (2)$$

$$[N \, \text{II}] \frac{j_{\lambda6548} + j_{\lambda6583}}{j_{\lambda5755}} = \frac{8.23 exp(2.50 \times 10^4/T)}{1 + 4.4 \times 10^{-3} n_e/\sqrt{T}} \quad (3)$$

$$[Ne \, \text{III}] \frac{j_{\lambda3869} + j_{\lambda3968}}{j_{\lambda3343}} = \frac{13.7 exp(4.30 \times 10^4/T)}{1 + 3.8 \times 10^{-5} n_e/\sqrt{T}} \quad (4)$$

$$[S \, \text{III}] \frac{j_{\lambda9532} + j_{\lambda9069}}{j_{\lambda6312}} = \frac{5.44 exp(2.28 \times 10^4/T)}{1 + 3.5 \times 10^{-4} n_e/\sqrt{T}} \quad (5)$$

j is the strength of the stated line, $n_e$ is the electron density and T the electron temperature. Figure 3 shows the graphical solution of the line ratio equations for [O III] (red), [N II] (green), [Ne III] (blue) and [S III] (magenta) for our 20kK (upper panel), 30kK (intermediate panel) and 40kK models (lower panel). The dashed line always depicts the derived temperature for the no-shock model whereas the solid line represents the temperature derived for the shock model. The electron temperatures are slightly higher for the shock model for the case of the 20kK model. Interestingly are the temperatures for the 30kK shock model slightly lower than for the no-shock model. This is to be understood in the context of the thermal balance in the two cases. In the 20kK case the extra heating due to the ionising radiation coming from the shock region causes the increase in the nebular gas temperature. In the case of the 30kK star however the increase in the heating by photoionisation is less in proportion (the star already radiates in this wavelength region), and this effect is overpowered slightly by the enhancement of some of the cooling emission lines, causing a slight net decrease in the nebular gas temperature. For the 40kK model, as expected, the differences between the no-shock and shock model are again almost negligible.
As expected, the line diagnostics of the electron density yielded no differences whatsoever between the no-shock and shock models. Using the graphical solution presented in Osterbrock & Ferland (2006) for the [O II]$\lambda3729/\lambda3726$ and the [S II]$\lambda6716/\lambda6731$ line ratio, we deduce an electron density of roughly $3700cm^{-3}$, $3400cm^{-3}$ and $2800cm^{-3}$ for our 20kK, 30kK and 40kK models, which are the values used in the electron temperature equations above.

### 3.4 The effect of shocks on strong lines in giant H II regions

An interesting question was to estimate the effect of shocks in stellar atmospheres on giant H II regions. The question was if the 20kK stars in a stellar cluster do significantly influence the overall ionising spectrum under the consideration of shocked stellar atmospheres.

We however found that the extra-emission produced by shocks in the atmospheres of 20kK stars is completely washed out by the emission of higher mass stars that photospherically emit in the same wavelength region. This effect would be only noticeable in a cluster where the largest star is indeed a 20kK star. Nevertheless we found in this context an interesting result concerning the He II $\lambda1640$ and He II $\lambda4686$ nebular lines. Table 2 shows only these two lines with a significant enhancement by shocks in the stellar atmosphere of a 30kK stars, all other lines show, compared to the 20kK model, already no differences. This is especially of interest since Stasińska & Izotov (2003) pointed out that the He II $\lambda4686$ nebular emission line of their sample of investigated H II galaxies occurred too frequently. The observed He II $\lambda4686$/H$\beta$ ratios could only be explained by assuming an additional X-ray radiation from a roughly $10^6$K hot thermal plasma. This finding goes perfectly along with the emission of high energy radiation produced by shocks in the atmospheres of the ionising star. Moreover, our derived shock structure results in maximum shocks in the stellar atmosphere temperatures of $1.1 \times 10^6 K$ for the 20kK model and $1.2 \times 10^6 K$ for the 30kK model, respectively, which is in almost perfect agreement with the finding of Stasińska & Izotov (2003). Nonetheless, despite the increase of the He II $\lambda4686$ line strength due to the extra emission of X-rays resulting from shock cooling zones, the enhancement is unfortunately much smaller than what is needed to explain the observations in the H II galaxies.

## 4 SUMMARY AND CONCLUSIONS

In this paper we have used the stellar atmosphere code WM-Basic (Pauldrach et al. 2001) and the photoionisation code MOCASSIN (Ercolano et al. 2003, 2005, 2008) to investigate whether or not the shock enhanced ionising flux from stellar atmospheres has a non-negligible influence on the emission line spectra of H II regions. A possible influence would lead to an impact on the determination of stellar and electron effective temperatures as well as electron densities. The question was if current stellar atmosphere models, which do not take shocks into account, and are used as inputs to the grid of photoionisation models in currently employed nebular emission line diagnostic diagrams would produce significant biases, and therefore whether a new generation of photoionisation model grids is required to solve the problem.

To answer this question we have run a grid of stellar atmosphere models using the WM-Basic code in shock and no-shock mode to generate input spectra for the MOCASSIN photoionisation code. The calculated WM-Basic grid ranged from 20kK to 50kK (60kK to 90kK) in steps of 10kK in the effective temperature for a surface gravity of log g = 3.9 (log g = 4.9) and solar abundances. The emergent ionising flux was then used to derive the corresponding number of ionising photons as well as the model's Strömgren radius. The derived quantities were used as an input in the MOCASSIN code. Based on these input parameters each MOCASSIN model yielded a corresponding ionisation parameter U. We used the $\eta_{S-Ne}$ parameter which is one of the most commonly employed method to estimate the mean $\overline{U}$ and $T_{eff}$ from ratios of strong nebular emission lines. The net effect of shocks is to simulate a star of higher effective temperature and thus a higher ionisation parameter in the nebular gas. Our results showed no change at all for the 40kK case. For cooler stars we indeed found small deviations for the 30kK non-shock case to the 30kK shock case. The effects on the 20kK case are dramatic, placing the 20kK shock model in a region of the diagnostic diagram (Fig. 2) expected to be populated by systems with roughly 30-35kK ionising stars and approximately a decade higher ionisation parameter U.

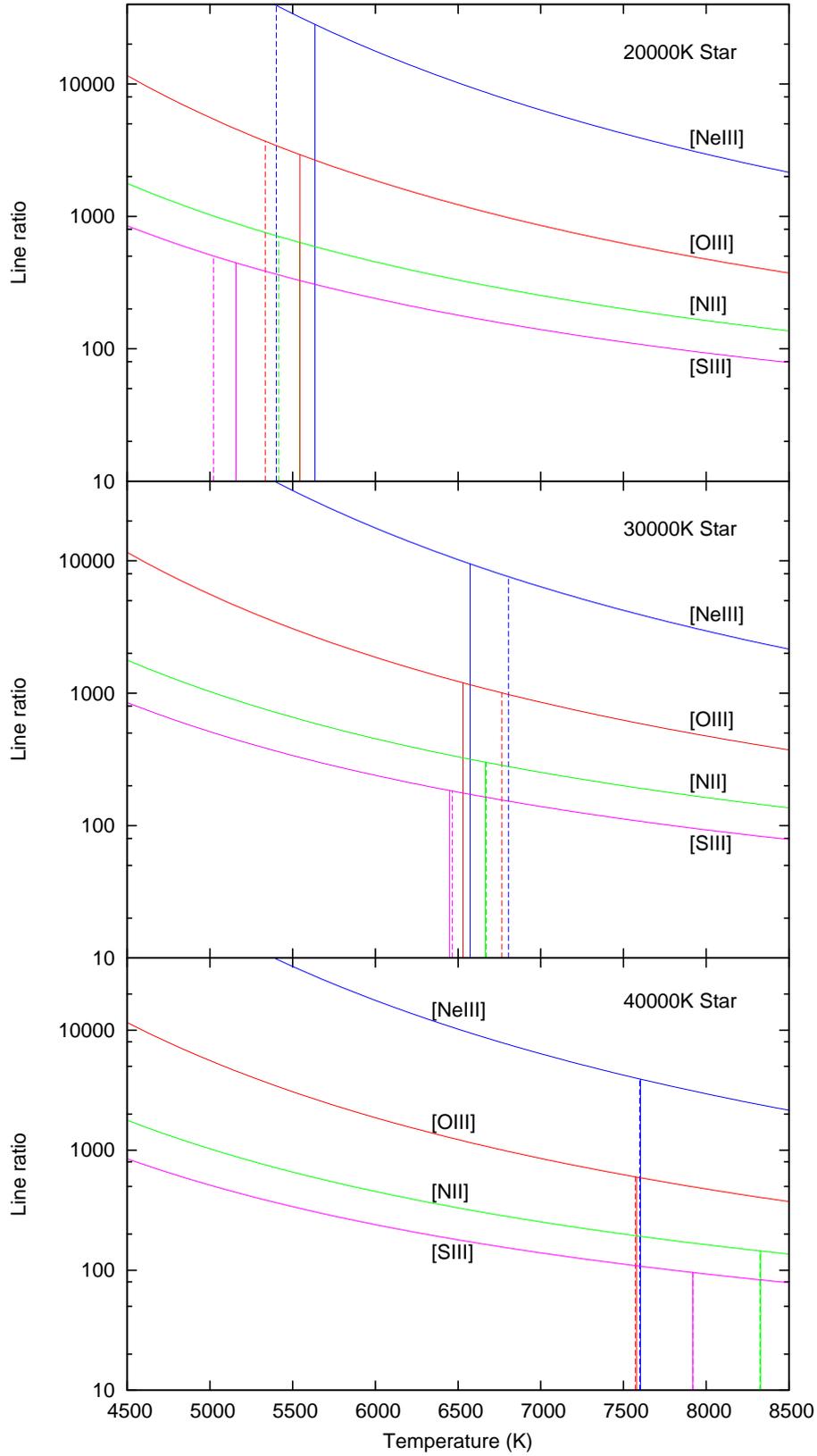

**Figure 3.** Graphical solution for the electron temperature line diagnostic equations (see Sec. 3.3) for the 20kK (upper panel), the 30kK (intermediate panel) and the 40kk model (lower panel). The calculated line ratios for the four investigated lines [O III] (red), [N II] (green), [Ne III] (blue) and [S III] (magenta) yield for the no-shock and shock model different temperatures. Each vertical dashed line points out the electron temperature obtained by the corresponding line ratio in case of the no-shock model whereas the vertical solid line points out the electron temperature obtained by the corresponding line ratio of the shock model.

| Line Ratio | 20 [10$^3$K] | | 30 [10$^3$K] | | 40 [10$^3$K] | |
|---|---|---|---|---|---|---|
| | Derived electron temperatures [K] | | | | | |
| | ns | s | ns | s | ns | s |
| [O ɪɪɪ] | 5334.72 | 5544.08 | 6764.84 | 6529.89 | 7571.73 | 7578.90 |
| [N ɪɪ] | 5415.73 | 5542.92 | 6671.09 | 6664.40 | 8324.86 | 8328.70 |
| [Ne ɪɪɪ] | 5401.27 | 5635.36 | 6805.02 | 6572.95 | 7597.79 | 7602.95 |
| [S ɪɪɪ] | 5022.46 | 5157.55 | 6465.15 | 6449.19 | 7919.35 | 7919.24 |

**Table 5.** Values obtained by the graphical solution of the line ratio equations (Eq. 2-5) for the no-shock and shock models of our 20kK, 30kK and 40kK models.

We also investigated a possible influence on the electron temperatures and densities for the no-shock and shock cases of our models. We used line ratio equations for the electron temperature and density determination as presented by Osterbrock & Ferland (2006). A graphical solution of the line ratios equations (see Fig. 3) yielded in the shock case for all temperatures values which are roughly 200K higher than for the 20kK no-shock model and showed again almost no differences between the no-shock and shock models in the 40kK case. Interestingly the obtained electron temperatures for the 30K case were all a bit lower in the shock case compared to the no-shock case. This is understood in the context of the thermal balance for each model. In the 20kK case the extra heating due to the ionising radiation coming from the shock region causes the increase in the nebular gas temperature. The 30kK star however already radiates in this wavelength region and therefore is the increase in the heating by photoionisation less in proportion. This effect is overpowered slightly by the enhancement of some of the cooling emission lines, causing a slight net decrease in the nebular gas temperature.

For the empirically derived electron densities we found no differences between the no-shock and shock models.

As an additional result we found our shock models to show an enhancement of the He ɪɪ λ4686 line. Stasińska & Izotov (2003) already observed an unusually frequent occurrence of the He ɪɪ λ4686 nebular emission line in their sample of investigated H ɪɪ galaxies. However, we find that the enhancement of the He ɪɪ λ4686 line is unfortunately much smaller than what would be needed to explain the observations of the H ɪɪ galaxies.

Overall we conclude that shock effects are only important for stellar sources with effective temperatures equal or lower than 30kK. The magnitude of the effect is also obviously dependent on the strength of the shock (which can be determined by modelling fits to the stellar emission lines, e.g. the *O* vɪ line, if observations are available for appropriate stellar objects) and thus might only show dramatic effects for a chosen sample of stellar sources and be likely unimportant for the majority of stellar sources.

**ACKNOWLEDGMENTS**

We would like to thank Nate Bastian for carefully reading the paper and giving helpful advice. This work was supported by the Deutsche Forschungsgemeinschaft (DFG) under grant Pa 477/4-1.